# A Bibliographic Study on Artificial Intelligence Research: Global Panorama and Indian Appearance

Amit Tiwari [1, 2], Susmita Bardhan [3], Vikas Kumar [1]

[1] DRTC, Indian Statistical Institute, Bangalore, Karnataka, India.

[2] DLIS, University of Calcutta, Kolkata, West Bengal, India.

[3] Rezoomex, Pune, India

**Abstract:**
The present study identifies and assesses the bibliographic trend in Artificial Intelligence (AI) research for the years 2015-2020 using the science mapping method of bibliometric study. The required data has been collected from the Scopus database. To make the collected data analysis-ready, essential data transformation was performed manually and with the help of a tool viz. OpenRefine. For determining the trend and performing the mapping techniques, top five open access and commercial journals of AI have been chosen based on their citescore driven ranking. The work includes 6880 articles published in the specified period for analysis. The trend is based on Country-wise publications, year-wise publications, topical terms in AI, top-cited articles, prominent authors, major institutions, involvement of industries in AI and Indian appearance. The results show that compared to open access journals; commercial journals have a higher citescore and number of articles published over the years. Additionally, IEEE is the prominent publisher which publishes 84% of the top-cited publications. Further, China and the United States are the major contributors to literature in the AI domain. The study reveals that neural networks and deep learning are the major topics included in top AI research publications. Recently, not only public institutions but also private bodies are investing their resources in AI research. The study also investigates the relative position of Indian researchers in terms of AI research. Present work helps in understanding the initial development, current stand and future direction of AI.

**Keywords:** Artificial intelligence, Bibliographic trend analysis, Topical trend, Keyword network analysis, Institutional collaboration, OpenRefine

## 1. Introduction

Artificial Intelligence (AI) is the concept of making machines, software and tools in such a way that it can understand and operate just like humans do. John McCarthy proposed the term "artificial intelligence" at the Dartmouth Conference in 1956, although its voyage began far earlier[1]. Alan Turing's work "Computing Machinery and Intelligence"[2], published in 1950, is considered as a benchmark for AI development[3]. According to John, AI is "the science and engineering of making intelligent machines, especially intelligent computer programs. It is related to the similar task of using computers to understand human intelligence, but AI does not have to confine itself to methods that are biologically observable"[4]. Clearly, AI was motivated by the human cognitive processes that include the components of human intelligence such as thinking, problem-solving, perceiving, learning, and decision making more efficiently, quickly and at a lower cost.

In a technical paper, 'Programs with common sense,' McCarthy[5] defined and wrote the first complete AI program. The program was pertinent to the narrower problems and failed to address the general problems[6]. The invention of perceptrons was another significant event in the evolution of AI[7]. Despite being criticized by experts, notably by Marvin Minsky[8], perceptron algorithms evolved strongly and are presently known as Neural networks[9][10]. Since its inception. The multidisciplinary approach brought an unprecedented change in AI research and practice[11]. Different domain experts incorporated interdisciplinary techniques such as hidden Markov models [12], stochastic search, signal processing[13], and optimization into AI thinking. As a result, a new area of machine learning emerged[6].

Recently, AI has become one of the most popular areas of research[11]. The exponential growth in AI research motivated us to visualize the bibliographic trend in AI. In this study, we have used Bibliographic factors for analyzing the trend in AI research.

The remaining part of the paper is laid out as follows: The review of literature is outlined in Section 2. The objectives, research question and the significance of the study are listed in Section 3. The work's research approach is explained in Section 4. The findings are listed in Section 5. In section 6, the findings are analyzed. Section 7 brings the work to a close by outlining future AI research strategies.

## 2. Literature review

Based on the datasets of Science Citation Index Expanded and Conference Proceedings Citation Index-Science, *Niu et al. (2016)* [14] conducted a spatially explicit bibliometric study for artificial intelligence research for the period 1990-2014. The work results include discussion on scholarly output, subject categories and prominent journals, author output and spatial location, international productivity and coordination, and current issues and research trends particular to AI. In a similar line, *Chung et al. (2017)*[15] evaluated AI thesis articles published by Korean Authors in the Science Citation Index Expanded journals during 1997-2016. In work, authors conducted frequency analysis and keyword network analysis in clusters of five years each and presented a bibliometric study. In another related study, *Chung et al. (2018)*[16] analyzed the artificial intelligence field based on patent data using keyword network analysis and blank technical analysis techniques. Furthermore, the work presents the direction of recent developments in AI. Using quantitative techniques *Park (2018)*[17] analyzed 758 cases of open patents of AI to evaluate the gradual development in AI technologies; the work also illustrates the trends in core AI technology.

Chassignol et al. *(2018)*[18] explored the educational panorama of AI. The study narrated the impact of artificial intelligence technologies on pedagogy and predicted the possible changes in the education landscape. Furthermore, the work describes the application of AI in decrypting student difficulties. In a different scope, Chong et al. *(2019)*[19] identified the practical trend of AI technology development by analyzing Open Source Software projects associated with AI. The work is based on the datasets, within the range of years 2000 to July 2018, available on

Github and analyzed by applying text mining techniques targeting topic information. The work indicates the characteristics of the collected projects and technical fields.

Similarly, *Vergeer (2020)*[20] examined the AI trend using newspaper articles published in the Netherlands between 2000 and 2018. The data was obtained from LexisNexis, a comprehensive news database. The result shows how reporting has been changing over time and how different kinds of newspapers wrote diversely about various topics in AI. In another related study, Castro *et al. (2019)*[21] presented a report based on China, the European Union, and the United States. The report examines six categories of criteria to determine the relative standing of the targeted regions in the AI economy: talent, research, development, adoption, data, and hardware. At the time of the report, It was stated that the US was the leader in the AI field. China was pacing up and the European Union was lagging. The study results found that despite having the highest AI publication, maximum data generation and adoption of AI technologies, China lagged in research funding.

Based on the literature, it was found that *Jiqiang Niu et al. (2016)* considered research articles for the study of trends in AI. Other literature were more focused on patents, software and thesis report to analyze data. Further, related studies by *Jiqiang Niu et al. (2016)* and *Myoung-Sug Chung et al. (2017)* are based on the web of science databases, namely the Science Citation Index Expanded. In contrast, the present work is based on the data collected from Scopus. As per our best knowledge, we did not find any study that measures the AI trends recently. AI practice is proliferating; therefore, the trend analysis will be a handy approach to understanding the subject's growth and direction. Thereby, the study of research trends in AI is significant.

**3. Objectives of the work**

The major objectives of the work are to

- Identify and demonstrate major research Institutes pursuing research in the field of AI based on top commercial and open access journal publications.
- Explore and analyze the topical trend of the statistical data of author keywords in the field of AI based on top commercial and open access journal publications of the domain.
- Identify the most prolific authors in AI and find out the most cited articles related to the AI domain and locate their geography-based trends.
- Explore the position of India in AI publications and identify the most productive Indian authors and institutions based on the publications in top commercial and open access AI journal publications.

The major contribution of this work is as follows:

- This study would provide a broader view of recent trends in Artificial Intelligence (AI).
- The work helps budding AI researchers to identify the universities and institutes where Artificial intelligence domains are flourishing.
- This paper also provides a keyword-based analysis that helps in identifying the topical trend in AI.

- The present work also enlists the top authors based on their number of publications and visualizes an author collaboration network for the selected journals that helps in identifying their productivity.

### 3.1 Research Questions

The study aimed at finding the answer to the following research questions:

**RQ1:** What are the major research Institutes pursuing research in AI for 2015-2020?

**RQ2:** What is the topical trend throughout 2015-2020 in AI?

**RQ3:** Who are the most frequently publishing authors in the field of AI?

**RQ4:** What are the most cited articles in the AI domain?

**RQ5:** What is the AI publishing trend across the globe?

**RQ6:** Where does India stand in AI publications and which Indian institutes and authors have maximum impact on AI publication?

### 4. Methodology

The present study is an exploratory research that has been conducted in the following stages: formulation of research question → data collection → data organization → data reporting and analysis (see Fig. 1).

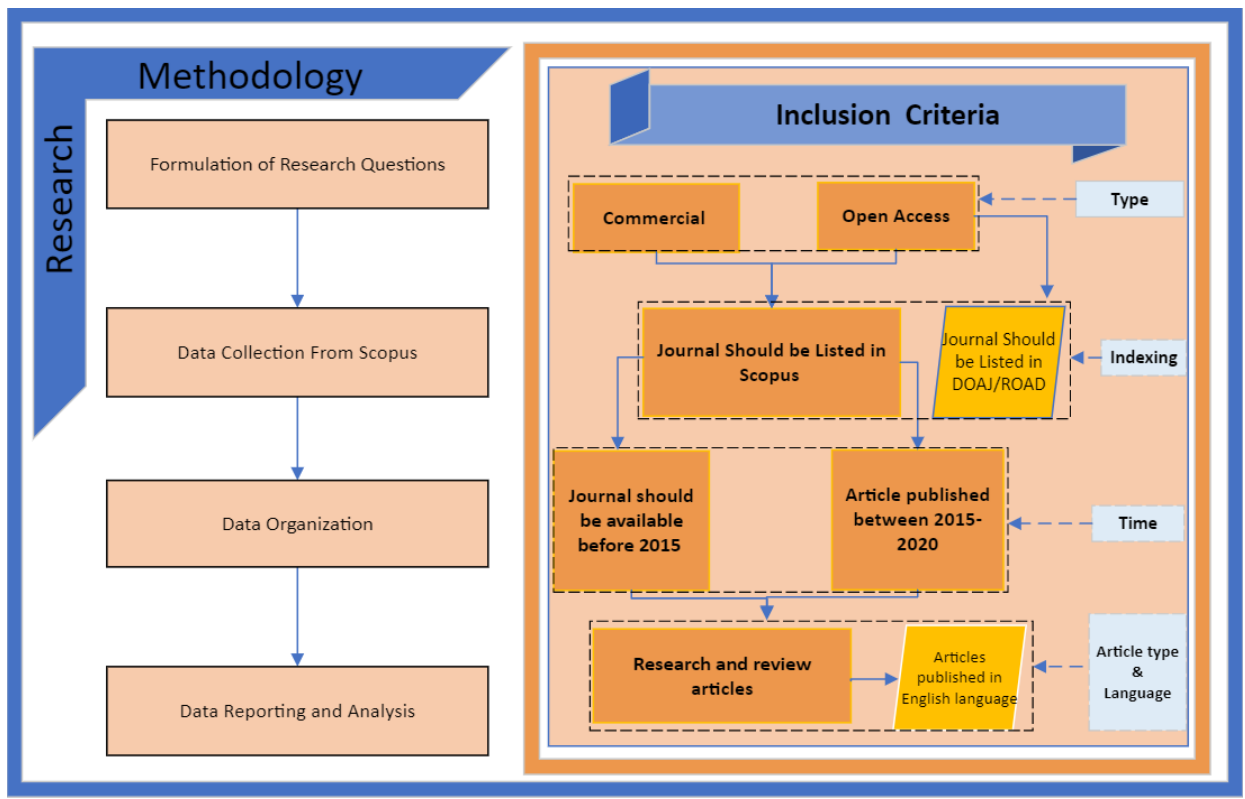

**Fig. 1: Research Methodology**

## 4.1 Research question formulation:

The research questions were identified based on the work objective and literature review. The literature review determined the gap in the trend-based bibliometric analysis of publications in AI journals.

**Table 1: Selected Journals for the study**

| Source title | CiteScore | Publisher | Listed in ROAD/ DOAJ | Open Access/Commercial |
|---|---|---|---|---|
| *IEEE Transactions on Pattern Analysis and Machine Intelligence* | 35.2 | IEEE | N/A | Commercial |
| *Foundations and Trends in Machine Learning* | 26.1 | Now Publishers Inc. | N/A | Commercial |
| *Physics of Life Reviews* | 20.9 | Elsevier | N/A | Commercial |
| *IEEE Transactions on Neural Networks and Learning Systems* | 17.6 | IEEE | N/A | Commercial |
| *IEEE Transactions on Fuzzy Systems* | 16.2 | IEEE | N/A | Commercial |
| *Journal of Machine Learning Research* | 9.3 | MIT Press | ROAD | Open Access |
| *Journal of Artificial Intelligence and Soft Computing Research* | 7.9 | Walter de Gruyter | DOAJ | Open Access |
| *Journal of Artificial Intelligence Research* | 6.4 | Elsevier | ROAD | Open Access |
| *ICT Express* | 5.9 | Korean Institute of Communications Information Sciences | ROAD | Open Access |
| *Computational Linguistics* | 4.8 | MIT Press | DOAJ | Open Access |

## 4.2 Data Collection:

For enquiring the research questions, data was obtained from Scopus. The study comprises five top-ranked commercial and open access AI journals each. The ranking was determined by the Scopus based citescore of the journals. The selected journals are listed in Table 1. Authors identified 2040 and 4840 articles from the selected open access and commercial journals respectively. For the study, authors exported the author name, author ID, the title of the articles, volume and issue numbers, affiliation, author name with affiliation information, author and indexed keywords of all 6880 papers from the Scopus database. We used Language, Publication

Period, Access, Screening, and document type parameters for the inclusion and exclusion of journals. We investigated both open access and subscription-based publications in our analysis. Only open access publications published in the ROAD (Directory of Open Access Resources) / DOAJ (Directory of Open Access Journals) and Scopus indexed top-rated journals were considered for this study. Furthermore, only those non-open access articles which are published in Scopus indexed top-ranked journals have been included in the study. In addition, only research and review articles published in English between 2015 and 2020 were considered for the study. Inclusion and exclusion criteria development is inspired by the work of Tiwari and Madalli (2021)[21] (see Fig. 1).

**4.3 Data Organization:**
For organizing the exported data we reviewed it, and found that some of the articles have messy or faulty information. In turn, it was essential to rectify the inconsistencies and resolve them. The rectification and data cleaning was performed both manually and with tool assistance. OpenRefine and Google spreadsheet were used for data cleaning.

**4.4 Data Reporting and Analysis:**
The cleaned data were tabulated using Google spreadsheets. Data analysis was performed both manually and with the help of tools such as OpenRefine, Spreadsheet and VOSViewer. Analysis was guided by the research objective. The study results and data analysis is reported in the relevant section of this article. Tool VOSviewer was used to derive the network for scientific collaboration analysis and co-word analysis.

**4.5 Result:**
The results of present work are categorized as article publishing trend; topical trend; authors publication trend; article productivity based on citation institutional; and country-wise productivity based on the publication. The detailed results are as follows:

**5. Analysis and Interpretation**

**5.1 Article publishing trend:**
Table 2 presents the number of articles published in the selected ten journals during 2015 - 2020. A total of 2040 and 4840 articles were retrieved from open access and commercial journals, respectively. From Table 2, the Journal of Machine learning research has the highest number of articles in open access whereas, IEEE Transactions on Neural Networks and Learning Systems has the highest number of articles in the commercial publisher category. Foundations and Trends in Machine Learning have published a total of 22 articles in this period, which is the least among the selected journals. In 2020, the number of articles published was highest, whereas, in 2017 least number of articles was published in the selected journals. The number of publications in open access journals is relatively lesser compared to commercial journals. The study results show a significant difference between the number of publications in the commercial and open access journals. One of the expected reasons could be the publication frequency; for example, "Journal of Machine Learning Research" is an open-access journal published eight times a year, whereas selected IEEE journals publish monthly.

**Table 2: Journal wise and year wise distribution**

| Journal Name | Number of articles published in | | | | | | |
|---|---|---|---|---|---|---|---|
| | 2015 | 2016 | 2017 | 2018 | 2019 | 2020 | Total |
| *IEEE Transactions on Pattern Analysis and Machine Intelligence* | 193 | 188 | 185 | 225 | 212 | 266 | **1269** |
| *Foundations and Trends in Machine Learning* | 3 | 3 | 3 | 4 | 5 | 4 | **22** |
| *Physics of Life Reviews* | 12 | 11 | 10 | 9 | 29 | 8 | **79** |
| *IEEE Transactions on Neural Networks and Learning Systems* | 274 | 224 | 255 | 535 | 318 | 560 | **2166** |
| *IEEE Transactions on Fuzzy Systems* | 181 | 132 | 139 | 315 | 192 | 345 | **1304** |
| *Journal of Machine Learning Research(OA)* | 139 | 232 | 149 | 169 | 183 | 250 | **1122** |
| *Journal of Artificial Intelligence and Soft Computing Research(OA)* | 20 | 19 | 20 | 16 | 22 | 20 | **117** |
| *Journal of Artificial Intelligence Research(OA)* | 47 | 65 | 67 | 69 | 61 | 64 | **373** |
| *ICT Express(OA)* | 29 | 40 | 40 | 37 | 54 | 92 | **292** |
| *Computational Linguistics(OA)* | 24 | 26 | 23 | 27 | 19 | 17 | **136** |
| ***Total no. of article*** | **922** | **940** | **890** | **1406** | **1095** | **1626** | **6880** |

**5.2 Institution-wise productivity:**

Fig. 2 ranks the top 10 institutions based on their number of publications in the top commercial and open access AI journals. The figure also illustrates the overall ranking of the institutes based on their publications during 2015-2020. From the result, it can be seen that the Chinese Academy of Sciences contributed to the most number of publications. The average yearly publication of the Chinese Academy of Science is more than 47 articles in the top AI journals. The remaining 9 institutes publish on an average of 16-22 articles per year. Further, CNRS contributed the most number of publications followed by the University of California at Berkeley in the open access journals. The institutions-wise productivity showcases that there is a significant difference between open access and commercial publications.

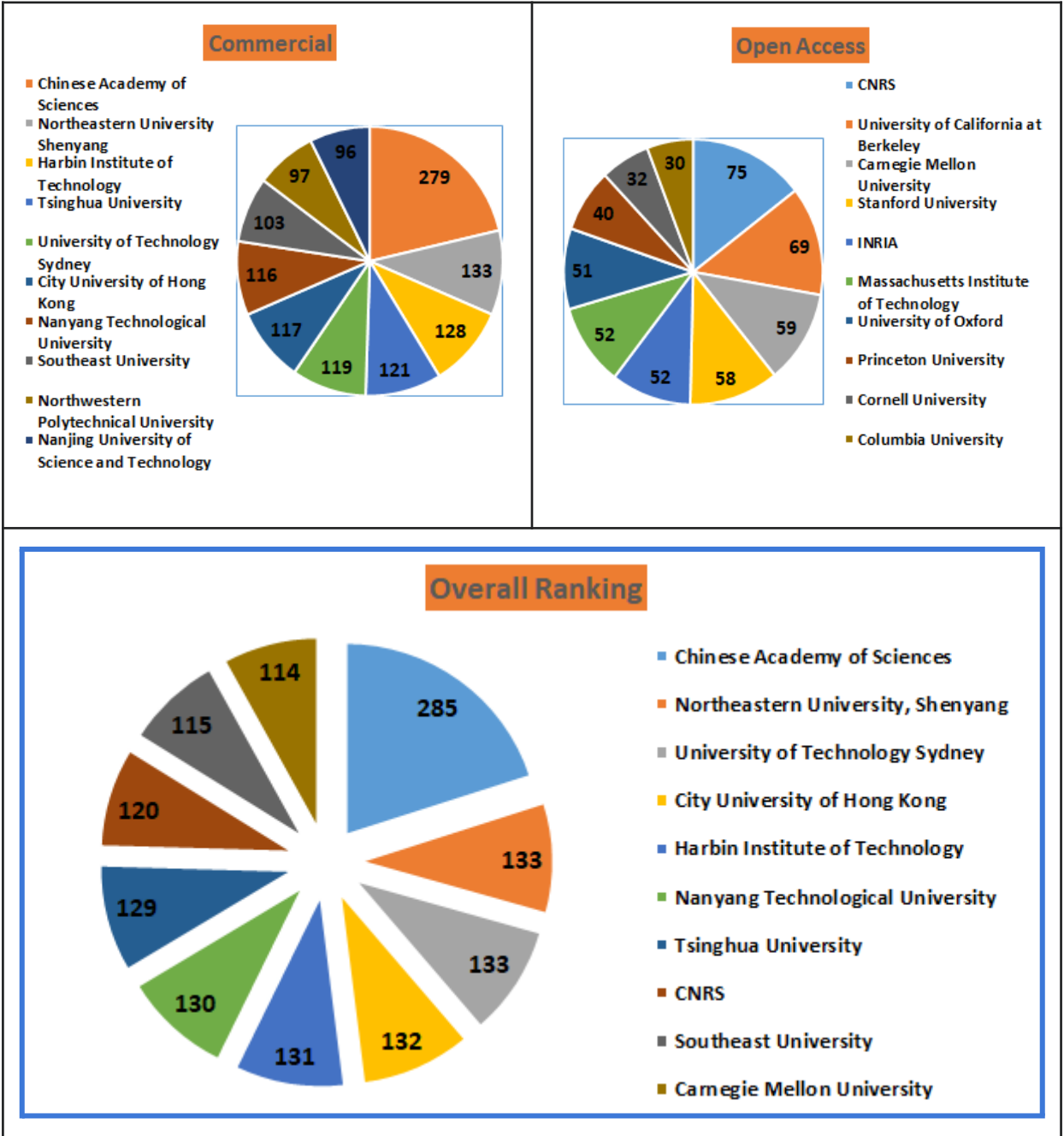

**Fig. 2: Institution-wise productivity**

### 5.3 Topical trend:

Fig. 3 illustrates the top twenty author-based keywords published during 2015-2020 in top AI journals. Neural networks and deep learning are the most frequently used keywords that reflect the area where the AI domain is heading.

Table 3: Top twenty keywords and their frequency during 2015-2020

| Keywords | 2015 | 2016 | 2017 | 2018 | 2019 | 2020 | Total |
|---|---|---|---|---|---|---|---|
| Neural Networks | 38 | 29 | 37 | 72 | 47 | 99 | 322 |
| Deep Learning | 6 | 21 | 18 | 49 | 52 | 113 | 259 |
| Convolutional Neural Networks | 3 | 7 | 10 | 42 | 42 | 55 | 159 |
| Machine Learning | 15 | 10 | 5 | 26 | 24 | 58 | 138 |
| Reinforcement Learning | 21 | 19 | 10 | 23 | 15 | 36 | 124 |
| Classification | 18 | 15 | 17 | 26 | 18 | 28 | 122 |
| Recurrent Neural Network | 9 | 12 | 24 | 25 | 8 | 24 | 102 |
| Support Vector Machine | 20 | 17 | 12 | 21 | 8 | 14 | 92 |
| Adaptive Control | 14 | 8 | 10 | 24 | 10 | 20 | 86 |
| Clustering | 7 | 15 | 10 | 16 | 16 | 21 | 85 |
| Non-linear Systems | 12 | 10 | 15 | 21 | 9 | 16 | 83 |
| Online Learning | 13 | 10 | 11 | 17 | 11 | 12 | 74 |
| Semi Supervised Learning | 13 | 7 | 9 | 13 | 11 | 20 | 73 |
| Feature Selection | 7 | 13 | 12 | 14 | 15 | 12 | 73 |
| Unsupervised Learning | 8 | 9 | 5 | 12 | 18 | 17 | 69 |
| Kernel Methods | 9 | 12 | 12 | 14 | 9 | 12 | 68 |
| Adaptive Fuzzy Control | 6 | 8 | 6 | 18 | 10 | 19 | 67 |
| Optimization | 5 | 9 | 6 | 6 | 16 | 23 | 65 |
| Supervised Learning | 9 | 6 | 12 | 10 | 15 | 11 | 63 |
| Transfer Learning | 4 | 6 | 7 | 18 | 11 | 17 | 63 |

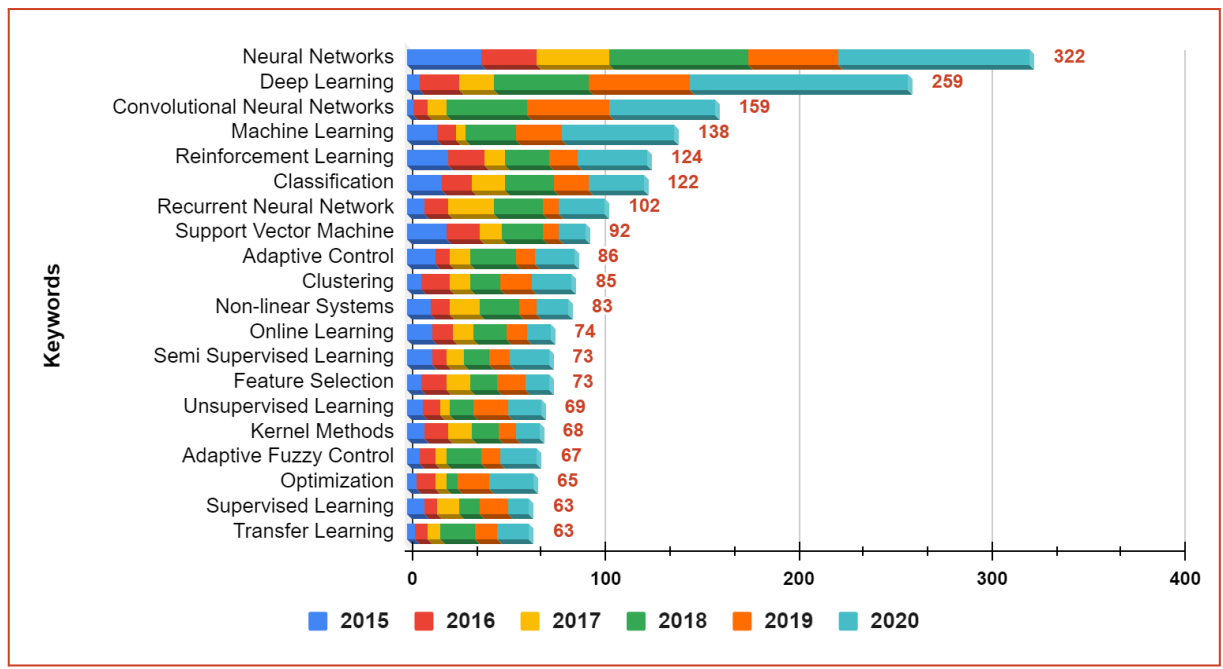

**Fig. 3: Topical Trend based on Author Keywords**

Deep learning has gained significant attention after 2015. The count of keywords, deep learning, increased around 18 times after its initial appearance. The author keyword, neural networks, has appeared most in the top AI journals in the last six years. From table 3 it can be observed that there is a constant change in the count of keywords over the years. That shows AI research and practice is witnessing newer concepts and technologies.

### 5.4 Authors with Article Frequency:

Table 4.1 enlists the top ten authors and the number of their publications in the selected commercial journals. In the Scopus database, there are many different authors whose names are the same. To maintain accuracy, we identified the publication counts based on the author IDs. From Table 4.1, after 2015 Li X. of Northwestern Polytechnic University, China, has published the most in the AI domain. The result shows that the authors have relatively preferred to publish in subscription-based journals, and therefore, it can be noted that the ranking of authors those published in commercial journals and overall (both open access and commercial) journals is the same.

Table 4.1: Top ten authors based on their publication in commercial journals

| S.no | Authors Name | Authors IDs | Affiliations | No of articles (Overall) | No of articles (in Commercial) |
|---|---|---|---|---|---|
| 1 | Li X. | 55936260100 | Northwestern Polytechnical University | 66 | 66 |
| 2 | Tao D. | 7102600334 | University of Sydney | 58 | 58 |
| 3 | Chen C.L.P. | 47861886300 | South China University of Technology | 52 | 52 |
| 4 | Shi P. | 36748941200 | University of Adelaide | 46 | 46 |
| 5 | Pedrycz W. | 56854903200 | University of Alberta | 45 | 45 |
| 6 | Cao J. | 7403354075 | Southeast University | 45 | 39 |
| 7 | Tong S. | 56496204400 | Liaoning University of Technology | 39 | 39 |
| 8 | Zhang H. | 56122851800 | Northeastern University, Shenyang | 37 | 37 |
| 9 | Huang T. | 8868586900 | Texas A&M University at Qatar | 37 | 37 |
| 10 | Yang G.H. | 7405751358 | Northeastern University, Shenyang | 35 | 35 |

Table 4.2: Top ten authors based on their publications in open access journals

| S.no | Authors Name | Authors IDs | Affiliations | No of articles |
|---|---|---|---|---|
| 1 | Wainwright M.J. | 7102128519 | University of California Berkeley | 14 |
| 2 | Schölkopf B. | 7004460308 | Max Planck Institute for Intelligent Systems | 9 |
| 3 | Mahoney M.W. | 7202006961 | University of California Berkeley | 9 |
| 4 | Liu H. | 26642968600 | Princeton University | 9 |
| 5 | Xing E.P. | 57203079532 | Carnegie Mellon University | 9 |
| 6 | Dunson D.B. | 7006472073 | Duke University | 8 |

| 7 | Bach F. | 7202286449 | INRIA, CNRS, PSL Research University, | 8 |
|---|---|---|---|---|
| 8 | Chen X. | 57192468738 | New York University | 8 |
| 9 | Michailidis G. | 6701747952 | University of Florida | 8 |
| 10 | Zhang T. | 7404373332 | Hong Kong University of Science and Technology | 8 |

From table 4.2, Wainwright M. J. of the University of California Berkeley has the credit to publish maximum articles in top-ranked open access AI journals. However, compared to subscription-based publications the publication in open access journals is significantly lesser.

**5.5 Citation:**

Tables 5.1 and 5.2 present the rank-wise citation data of top ten articles published in commercial and open access journals respectively. Most of the articles having the highest citation belong to *IEEE Transactions on Pattern Analysis and Machine Intelligence,* followed by the Journal of Machine Learning Research. The work *Faster R-CNN: Towards Real-Time Object Detection with Region Proposal Networks* got the highest citation score of 3686. The article was published in 2017 and written by Ren S., He K., Girshick R., Sun J.

**Table 5.1: Most cited articles published in commercial journals**

| Title | Publication Year | Authors | Cited by | Journal Name |
|---|---|---|---|---|
| Faster R-CNN: Towards Real-Time Object Detection with Region Proposal Networks | 2017 | Ren S., He K., Girshick R., Sun J. | 4837 | IEEE Transactions on Pattern Analysis and Machine Intelligence |
| DeepLab: Semantic Image Segmentation with Deep Convolutional Nets, Atrous Convolution, and Fully Connected CRFs | 2018 | Chen L.-C., Papandreou G., Kokkinos I., Murphy K., Yuille A.L. | 3230 | IEEE Transactions on Pattern Analysis and Machine Intelligence |
| SegNet: A Deep Convolutional Encoder-Decoder Architecture for Image Segmentation | 2017 | Badrinarayanan V., Kendall A., Cipolla R. | 3223 | IEEE Transactions on Pattern Analysis and Machine Intelligence |
| High-speed tracking with kernelized correlation filters | 2015 | Henriques J.F., Caseiro R., Martins P., Batista J. | 2932 | IEEE Transactions on Pattern Analysis and Machine Intelligence |
| Image Super-Resolution Using Deep Convolutional Networks | 2016 | Dong C., Loy C.C., He K., Tang X. | 2737 | IEEE Transactions on Pattern Analysis and Machine Intelligence |
| Fully Convolutional Networks for Semantic Segmentation | 2017 | Shelhamer E., Long J., Darrell T. | 2320 | IEEE Transactions on Pattern Analysis and Machine Intelligence |

| Title | Year of Publication | Authors | Cited by | Journal Name |
|---|---|---|---|---|
| Spatial Pyramid Pooling in Deep Convolutional Networks for Visual Recognition | 2015 | He K., Zhang X., Ren S., Sun J. | 2128 | IEEE Transactions on Pattern Analysis and Machine Intelligence |
| Object tracking benchmark | 2015 | Wu Y., Lim J., Yang M.-H. | 1694 | IEEE Transactions on Pattern Analysis and Machine Intelligence |
| Global contrast based salient region detection | 2015 | Cheng M.-M., Mitra N.J., Huang X., Torr P.H.S., Hu S.-M. | 1403 | IEEE Transactions on Pattern Analysis and Machine Intelligence |
| LSTM: A Search Space Odyssey | 2017 | Greff K., Srivastava R.K., Koutnik J., Steunebrink B.R., Schmidhuber J. | 1292 | IEEE Transactions on Neural Networks and Learning Systems |

**Table 5.2: Most cited articles published in open access journals**

| Title | Year of Publication | Authors | Cited by | Journal Name |
|---|---|---|---|---|
| Domain-adversarial training of neural networks | 2016 | Ganin Y., Ustinova E., Ajakan H., Germain P., Larochelle H., Laviolette F., Marchand M., Lempitsky V. | 1319 | Journal of Machine Learning Research |
| Accelerating t-SNE using tree-based algorithms | 2015 | Van Der Maaten L. | 828 | Journal of Machine Learning Research |
| End-to-end training of deep visuomotor policies | 2016 | Levine S., Finn C., Darrell T., Abbeel P. | 799 | Journal of Machine Learning Research |
| MLlib: Machine learning in Apache Spark | 2016 | Meng X., Bradley J., Yavuz B., Sparks E., Venkataraman S., Liu D., Freeman J., Tsai D.B., Amde M., Owen S., Xin D., Xin R., Franklin M.J., Zadeh R., Zaharia M., Talwalkar A. | 730 | Journal of Machine Learning Research |
| Stereo matching by training a convolutional neural network to compare image patches | 2016 | Zbontar J., Lecun Y. | 579 | Journal of Machine Learning Research |
| Deeply-supervised nets | 2015 | Lee C.-Y., Xie S., Gallagher P.W., Zhang Z., Tu Z. | 497 | Journal of Machine Learning Research |
| Imbalanced-learn: A python toolbox to tackle the curse of imbalanced datasets in machine learning | 2017 | Lemaitre G., Nogueira F., Aridas C.K. | 391 | Journal of Machine Learning Research |

| CVXPY: A Python-embedded modeling language for convex optimization | 2016 | Diamond S., Boyd S. | 320 | Journal of Machine Learning Research |
| A survey on LPWA technology: LoRa and NB-IoT | 2017 | Sinha R.S., Wei Y., Hwang S.-H. | 307 | ICT Express |
| A primer on neural network models for natural language processing | 2016 | Goldberg Y. | 287 | Journal of Artificial Intelligence Research |

From Table 5.1 and 5.2, it can be noted that top 9 cited articles are published in commercial journals. Thereby, the top-cited open access article *Domain-adversarial training of neural networks* published in the *Journal of Machine Learning Research* gets overall rank of 10.

### 5.6 Country Wise AI Publication:

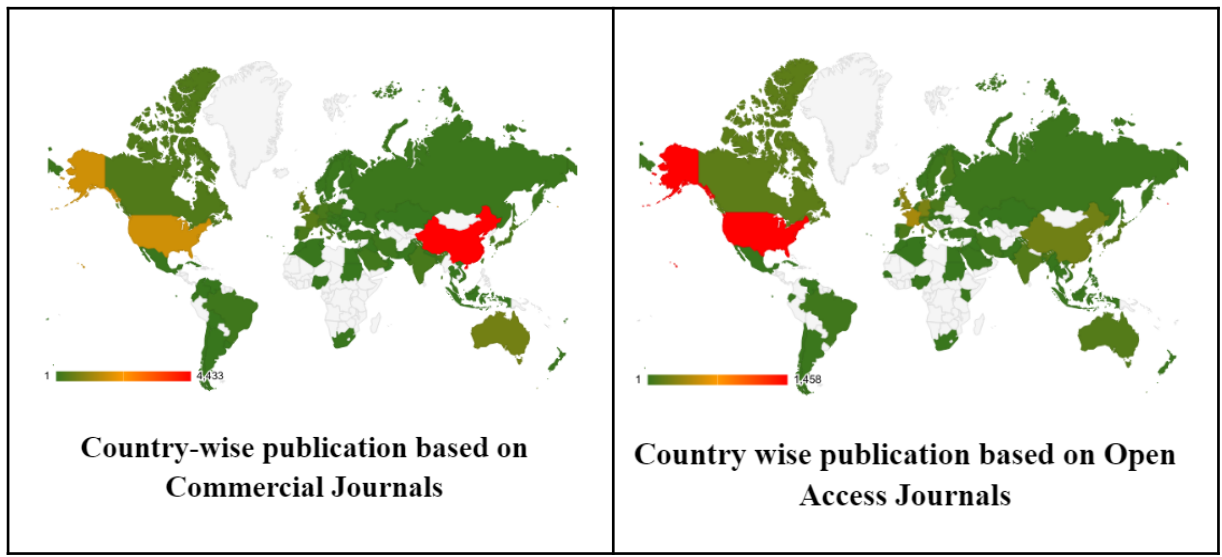

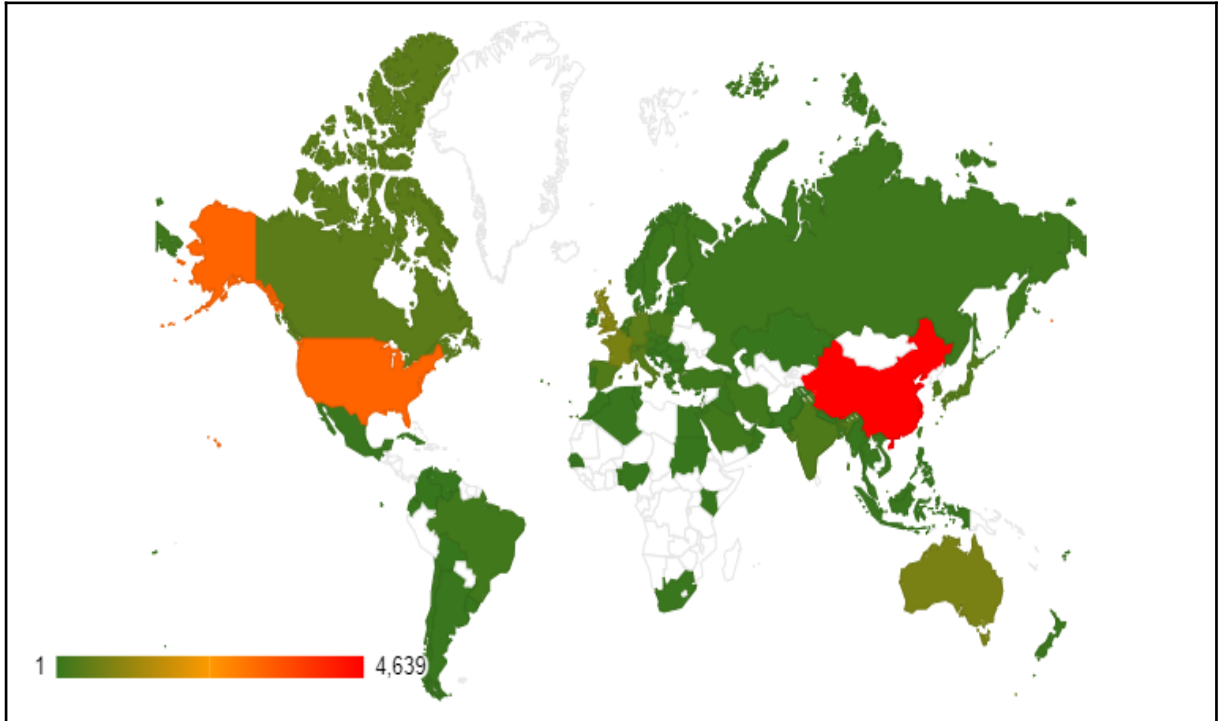

**Fig. 4: Country-wise AI publication**

Country-wise distribution of AI research and development is depicted in Fig. 4. The scholars affiliated to countries China and US-based institutions have the credit to publish most of the review and research articles in the domain of AI. China and US-based institutions had contributed to 4639 and 3124 articles, respectively among the selected journals. The United Kingdom is the following country on the list. The scholars from UK-based institutions have contributed to the 890 articles. The map illustrates the uneven distribution of AI research contributions. We can identify that African countries have significantly lesser publications in the top AI journals from the graphical representation. Among commercial journals, China with 4433, the USA with 1666 and Australia with 656 are ranked first, second and third respectively. We can see drastic change i.e. approximately 2.5 times difference among top 3 ranked publishing countries. Among open access journals, the United States, France and the United Kingdom are the top 3 countries with 1458, 406 and 305 articles respectively.

**5.7 Industries in AI Research:**

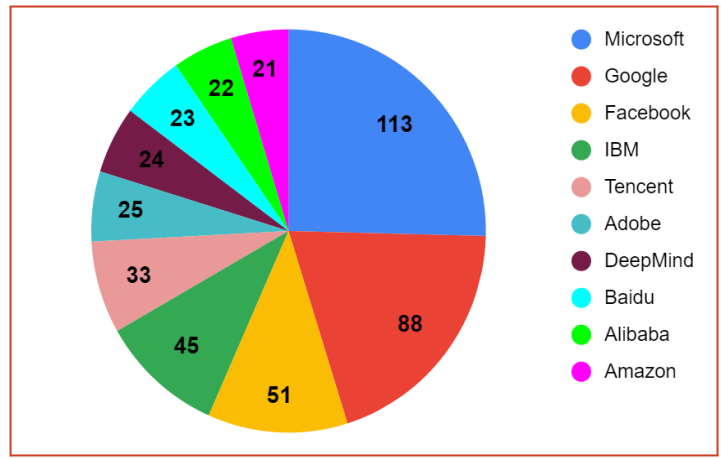

**Fig. 5: Publications by Industries in AI Research**

Artificial Intelligence is a practice-oriented field of research and development (R&D). Therefore, a significant number of industries are involved in R&D of AI. Microsoft, Google, Facebook, IBM and Tencent are the leading players in R&D of AI. Fig 5 illustrates the publication trends in the area of AI. Among industries, Microsoft has top publications, followed by Google, Facebook, IBM and others.

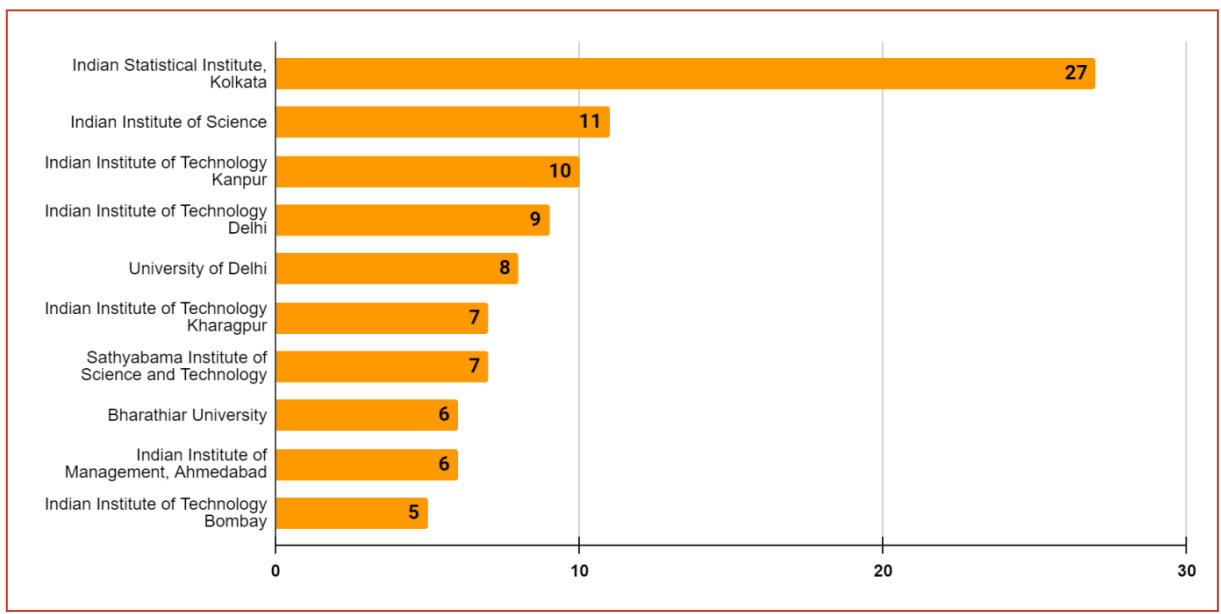

**Fig. 6: Top Indian Institutes based on their number of publications in AI journals from 2015 to 2020**

Based on the number of articles published in top AI journals from 2015 to 2020, India's place in the world ranking is 15th. Fig. 6 depicts that the Indian Statistical Institute, Kolkata, has contributed the most articles. Indian Institute of Science, Bangalore, has been ranked second. In the list, six institutes are from the engineering and technology category, two from universities (Delhi University and Bharathiar University) and two are special autonomous institutes (Indian Statistical Institute & Indian Institute of Science).

**5.9 Indian Authors:**

**Table 6: Top 10 Indian authors based on publications in top AI journals**

| Indian Authors | Institute | Occurrences |
|---|---|---|
| Pal N.R. | Indian Statistical Institute | 12 |
| Aggarwal M. | Indian Institute of Management, Ahmedabad | 6 |
| Gupta P. | University of Delhi | 5 |
| Rakkiyappan R. | Bharathiar University | 5 |
| Muhuri, P.K. | South Asian University | 4 |
| Verma, N.K. | Indian Institute of Technology, Kanpur | 4 |
| Das S. | Indian Institute of Technology, Patna | 3 |
| Pal, S.K., | Indian Statistical Institute, Kolkata | 3 |
| Behera L. | Indian Institute of Technology, Kanpur | 3 |
| Mehlawat, M.K. | University of Delhi | 3 |
| Sawhney, R.S. | Guru Nanak Dev University | 3 |

Table 6 includes the top 10 Indian authors having maximum publications in the selected journals. Among the Indian authors N.R. Pal of Indian Statistical Institute, Kolkata, has foremost publications.

## 6. Discussion

The present study is based on secondary data i.e., taken from Scopus. During data collection we found that for some of the cases, data was incomplete and inconsistent, e.g., the keywords were missing for some of the articles. In such cases, authors read the abstract and title of the work and derived the keywords. To the best of our knowledge, we have made the data consistent and complete before analysis. Additionally, for some of the parameters (such as topical terms, Industries and Indian context) the available open access data is almost insignificant or a combined analysis of open access and commercial journals make better sense. Therefore, for such parameters, we have performed the analysis combining the open access and commercial bibliographic information.

The study results contain the bibliographic information of the top five open access and commercial journals of domain Artificial Intelligence. As depicted in Table 2 there is a substantial difference between the numbers of publications. It was identified that, on average, selected open access journals take comparatively more time to publish their article. Moreover, the publication frequency of open access journals is also relatively lesser. It is noticed, there is significant growth in the published literature in both open (71%) and commercial journals (78%). However, the percentage growth in the literature of commercial journals is relatively more.

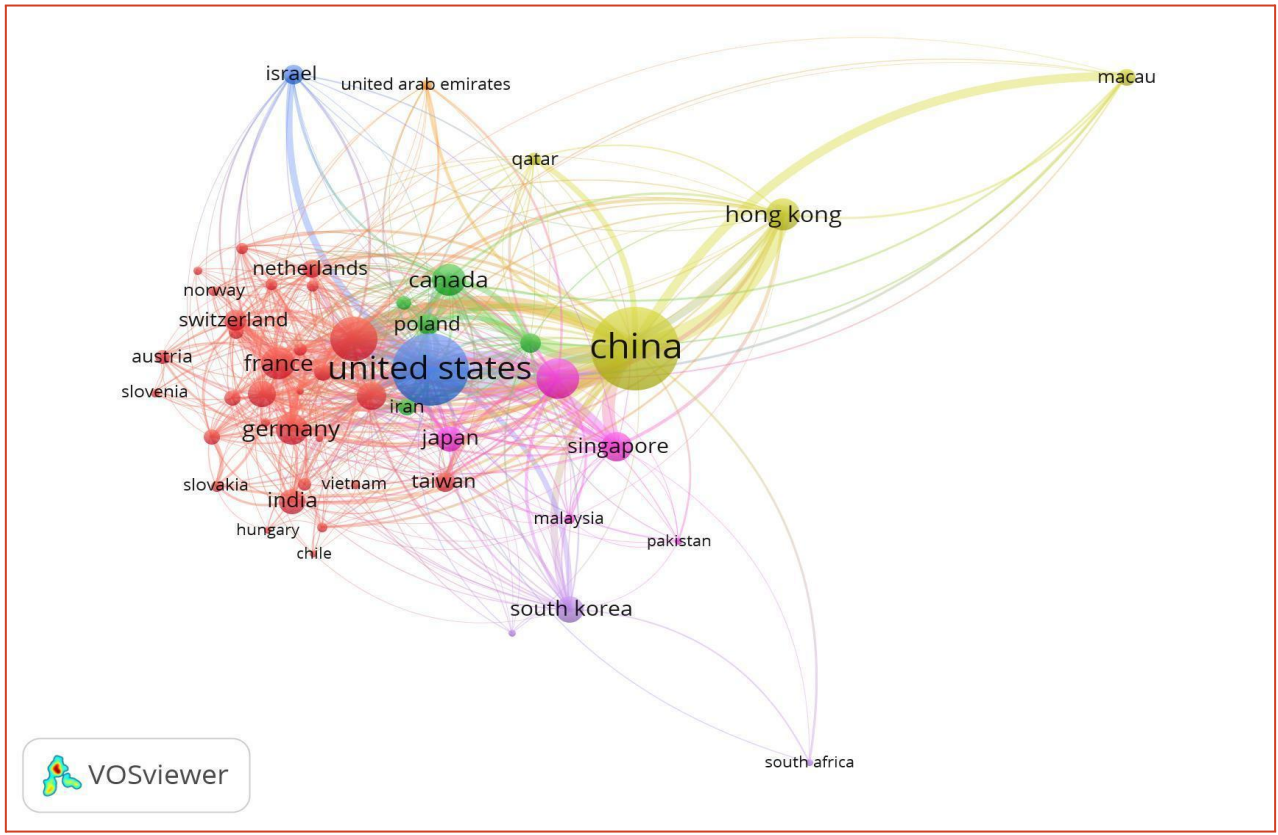

**Fig. 7: Country-wise co-authorship pattern**

Figure 2 shows that the majority of AI research published in leading AI journals comes from China-based institutes. Top 10 institutions published a total of 1422 documents. Combinedly for open access and commercial journals, it was noted that more than 29% of top authors are from China, 20% of authors are from the USA while 6% of authors are from the UK. When it comes to major commercial journals, China and the United States have a 38.5% and 14.5% stake, respectively. In contrast, in the open access scenario, China publishes 4.7% of articles, while United States based authors have a total of 34.5% of publications.

Co-authorship patterns among the countries and institutions are shown in Figures 7 and 9. From the figure, a notable fact is a strong pattern in co-authorship between the authors of China-UK, China-Hong Kong, China-Macau and China-Australia. In light of this fact, it can be noticed that while China is working with the scholars of several other countries, the USA lacks it significantly.

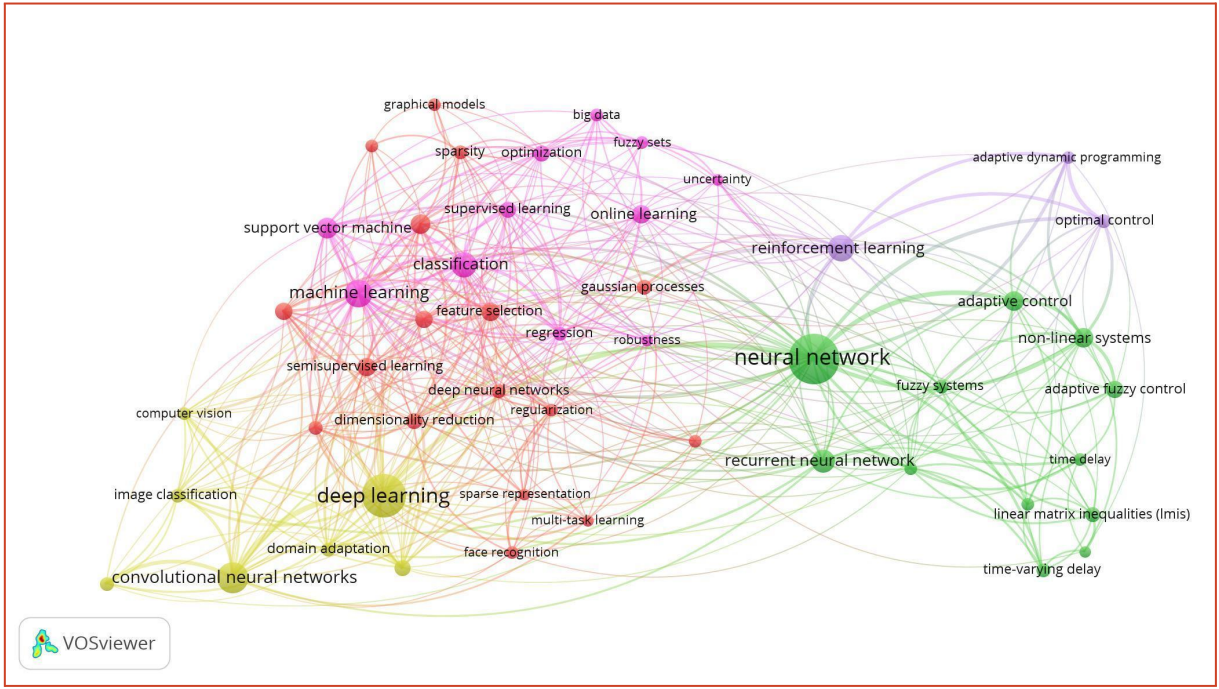

**Fig. 8: Co-Occurrence of author keywords**

The topical trend shows that new techniques and technologies are being adopted to improve AI practice. Deep learning, neural networks, convolutional neural networks, machine learning and classification are adopted in many research works. Keyword analysis indicates that the tools and techniques related to machine learning and the deep learning algorithm occurred the most number of times. Recently, the neural network, which is a part of deep learning, emerged as the most used and prominent technique in AI. Figure 8 illustrates the co-occurrence of author keywords.

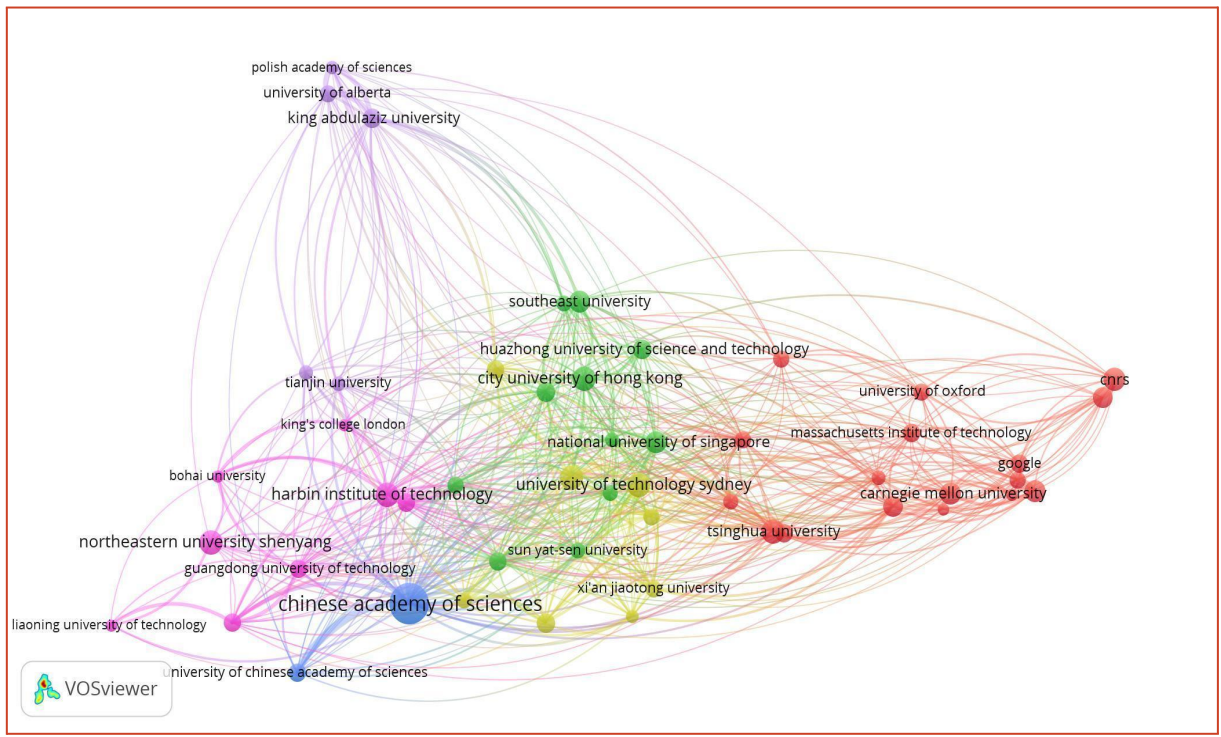

**Fig. 9: Institution wise co-authorship pattern**

AI research and development works are primarily conducted in China and the USA. The research works can be linked with the research funding and encouragement of the respective government and private stakeholders[22]. In this regard, many universities in China are devoted to a considerable amount of resources in science and technology, for example, the funding agency Chinese Academy of Science has more than 50 laboratories dedicated to areas of computer science and artificial intelligence and apparently, from Fig. 2, it has 15% share over the top publication.

Authors of the most cited articles, in most instances, belong to China, the United States and the United Kingdom. The study shows that the top 15 most cited papers are published in only three journals: Journal of Machine Learning Research, IEEE Transactions on Neural Networks and Learning Systems, IEEE Transactions on Pattern Analysis and Machine Intelligence. Further, it was noticed that around 67% of the top-cited articles are published in IEEE Transactions on Pattern Analysis and Machine Intelligence. Also, it was critical that the top cited articles of the same journal received 25483 citations (84%).

Though in AI publications, India stands at 15th rank, its overall AI publication is minimal i.e., 1.68% of the total publications in the top AI journals. The rank of India in open access scenario is 11th whereas, it stands at rank 14th in terms of publications in commercial journals. Indian institutions contributing to AI publishing are funded mainly by the government and dedicated to science and technology. At the national level, the Computer and Communication Sciences Division of the Indian Statistical Institute have published the most articles on AI.

## 7. Conclusion

AI is rapidly growing and it is not bound to a particular domain but ubiquitous. The scope of AI is broadened into the purview of sciences, social sciences and humanities. Though the researchers, corporates and governments are heavily engaged in AI research and development worldwide, the research trend shows its peak is yet to come. Present work returns all the determined research questions, particularly identifying topical trends, major research institutes, most frequently publishing authors, most cited articles in the global publishing trend, and Indian stance concerning AI publication. Based on data, this paper deduces that China, the United States and the United Kingdom are the key players in AI. The publications in commercial journals are significantly more than the open access journals.

Moreover, the articles published in commercial journals, specifically IEEE, have received more citations. The collaboration pattern and citation data indicate the advantages of inter institute research. Further, it was noticed that the open access journals lack in attracting the scholars to publish. Therefore, the open-access publishers should modify their existing strategies to attract more viewers, researchers and authors.